\documentclass[12pt]{article}
\usepackage{cite,graphicx,amssymb,amsfonts,hangcaption,color}

\renewcommand{\theequation}
{\arabic{section}.\arabic{equation}}

\makeatletter
\@addtoreset{equation}{section}
\@addtoreset{footnote}{section}
\makeatother

\makeatletter
\renewcommand\appendix{\par
\setcounter{section}{0}%
\setcounter{subsection}{0}%
\gdef\thesection{Appendix \@Alph\c@section }
\renewcommand{\theequation}
{\Alph{section}.\arabic{equation}}
}
\makeatother

\makeatletter
\def\eqnarray{ \stepcounter{equation} \let\@currentlabel=\theequation
\global\@eqnswtrue
\global\@eqcnt\z@
\tabskip\@centering
\let\\=\@eqncr
$$\halign to \displaywidth\bgroup\@eqnsel\hskip\@centering
$\displaystyle\tabskip\z@{##}$&\global\@eqcnt\@ne
\hfil$\displaystyle{{}##{}}$\hfil
&\global\@eqcnt\tw@$\displaystyle\tabskip\z@{##}$\hfil
\tabskip\@centering&\llap{##}\tabskip\z@\cr}
\makeatother

\makeatletter
\def\@arrayacol{\edef\@preamble{\@preamble \hskip .5\arraycolsep}}
\def\array{\let\@acol\@arrayacol \let\@classz\@arrayclassz
\let\@classiv\@arrayclassiv \let\\\@arraycr\def\@halignto{}\@tabarray}
\makeatother

\renewcommand{\arraystretch}{1.6}

\makeatletter
\newcounter{subeqncnt}
\def\thesubeqncnt{\alph{subeqncnt}}
\def\subequations{\begingroup%
\stepcounter{equation}\edef\@tempa{\theequation}%
\let\c@equation\c@subeqncnt\c@subeqncnt\z@
\edef\theequation{\@tempa\noexpand\thesubeqncnt}}

\makeatother

\newcommand{\captionfonts}{\small}
\makeatletter 
\long\def\@makecaption#1#2{%
\vskip\abovecaptionskip
\sbox\@tempboxa{{\captionfonts #1: #2}}%
\ifdim \wd\@tempboxa >\hsize
{\captionfonts #1: #2\par}
\else
\hbox to\hsize{\hfil\box\@tempboxa\hfil}%
\fi
\vskip\belowcaptionskip}
\makeatother 

\newcommand{\dd}{{\rm d}}

\begin{document}

\setlength{\baselineskip}{7mm}
\begin{titlepage}
\begin{flushright}
{\tt APCTP-Pre2009-011} \\
{\tt arXiv:0910.3722[hep-th]} \\
April 2010
\end{flushright}

\vspace{1cm}

\begin{center}

{\Large 
Magnetic Conductivity and Chern-Simons Term \\
in Holographic Hydrodynamics of Charged AdS Black Hole
}

\vspace{1cm}

{\sc{Yoshinori Matsuo}}$^\dagger$\footnote{
Present address: 
{\it Harish-Chandra Research Institute, 
Chhatnag Road, Jhusi, Allahabad 211019, India} \\
{\sf ymatsuo@hri.res.in}
}, \ 
{\sc{Sang-Jin Sin}}$^\ddagger$, \\
{\sc{Shingo Takeuchi}}$^\dagger$
\ and \
{\sc{Takuya Tsukioka}}$^\dagger$

\vspace*{3mm}

$\dagger$
{\it{Asia Pacific Center for Theoretical Physics},} \\
{\it{Pohang, Gyeongbuk 790-784, Korea}} \\
{\sf{ymatsuo, shingo, tsukioka@apctp.org}}
\\
$\ddagger$
{\it{Department of Physics,}}
{\it{Hanyang University,}}
{\it{Seoul 133-791, Korea}} \\
{\sf{sjsin@hanyang.ac.kr}}

\end{center}

\vspace{1cm}

\begin{abstract}
We study the effects of the Chern-Simons term in the hydrodynamics
of the five-dimensional Reissner-Nordstr\"om-AdS background.
We work out the decoupling problem of the equations of motion 
and calculate the retarded Green functions explicitly. 
We then find that the Chern-Simons term induces the magnetic conductivity 
caused by the anomaly effect. It is increasing function of temperature 
running from a non-zero value at zero temperature to 
  the twice the value at infinite temperature.    
\end{abstract}

\end{titlepage}

\section{Introduction}

Much efforts to use AdS/CFT correspondence~\cite{ads/cft, gkp, w}
to understand the strongly coupled field theory
have been performed after the discovery of low viscosity
in the gravity dual~\cite{pss0}.
The way to include a chemical potential was figured out in
the context of probe brane
embedding~\cite{ksz,ht,nssy1,kmmmt,nssy2,bergman,ubc,n}.
Even an attempt to map the entire process of RHIC experiment
in terms of the gravity dual~\cite{ssz} was made.
In spite of differences in QCD and the supersymmetric gauge theories,
it is expected that some of the properties are shared based on
the universality of low energy physics. In this respect,
the hydrodynamics is particularly interesting.
The calculation scheme for transport coefficients is to use Kubo
formula, which gives a relation to the low energy limit of Wightman
Green functions.
In AdS/CFT correspondence, one calculates the retarded Green
function which is related to the Wightman function
by fluctuation-dissipation theorem.
Such scheme has been developed in a series of
papers~\cite{ss,pss,pss2,hs,ks}.

In our previous papers~\cite{s,gmsst,gmssty,gmsst2},
we used the five-dimensional Reissner-Nordstr{\"o}m-Anti-deSitter
background~(RN-AdS$_5$)~\cite{mas,ss2,mno, saremi, bbn} to model the
dense baryonic media.
However, here the $U(1)$ charge in RN-AdS$_5$ black hole background is 
the $R$-charge and the Chern-Simon term describes the anomaly effect.
RN-AdS$_5$ can be also obtained from the STU solution~\cite{cvetic} by
taking its diagonal-part, and the Chern-Simons term is essential 
since the $U(1)$ is anomalous unlike the baryon $U(1)$ charge case.

It is generally known that the Chern-Simons term has relations
with the parity symmetry breaking and the triangle anomaly,
and its importance was pointed out in~\cite{Minwalla}.
It was shown that DC conductivities appear as coefficients
concerning pseudo-vectors and pseudo-tensors
in the derivative expansion of energy momentum tensor and conserved current,
when the background is boosted and the charge and mass vary slowly
with the spacetime coordinates~\cite{yarom, india, son1}.
As for the Chern-Simons term in the RN-AdS$_5$ background theory,
new understanding has been obtained recently~\cite{Son:2009tf}.
In their study, from the fact that the Chern-Simons term has to
do with the triangle anomaly, a relation between its coefficient
and the vorticity-induced term has been newly revealed.
Studies with the Chern-Simons term to obtain a description 
for the real QCD plasma in RHIC have been performed 
in~\cite{Torabian:2009qk,Yee:2009vw}.
In \cite{Yee:2009vw}, the time-dependent chiral magnetic conductivity 
has been calculated,
in which the property of the five-dimensional Chern-Simons term
corresponding to the four-dimensional anomaly plays an important role.
On the other hand, in~\cite{Torabian:2009qk}, 
exploiting multiple $U(1)^3$ symmetries in the STU model,
the viscous hydrodynamics of hot conformal field theory plasma
with $U(1)^{N_f}$ flavor symmetry as well as $SU(2)_{\rm{I}}$ 
non-Abelian iso-spin symmetry has been studied.

The aim of this paper is to complete the leading order hydrodynamics
in the equations of motion for RN-AdS$_5$ with the Chern-Simons term.
Notice that the presence of the Chern-Simons term does not change
the background spacetime, but it can change the perturbative
spectrum around the solution.
We calculate the retarded two-point Green functions explicitly and get
the transport coefficients in the presence of Chern-Simons term.
 
This paper is organized as follows:
In section~\ref{sec:Setup}, we describe the regularized action
with the Chern-Simons term and the RN-AdS$_5$ background
obtained from the STU model in the equal three charges.
The perturbations from the background are solved
under the hydrodynamic approximation
in section~\ref{sec:Perturb}.
In this procedure, we perform the decoupling procedure
by using the master variables.
In section~\ref{sec:dGQP},
we obtain the two-point retarded Green functions in the dual
Quark-Gluon-Plasma (QGP) via AdS/CFT correspondence.
We then investigate the effect of the Chern-Simons term in the bulk
gravity to the hydrodynamics of the dual QGP,
and summarize this study in section~\ref{sec:summary}.
In \ref{app:Green},
we give the procedure to calculate the retarded Green function
of the boundary theory in AdS/CFT correspondence formulated
by Son and Starinets.

\section{Setup in the dual gravity}\label{sec:Setup}

We first consider the STU solution~\cite{cvetic}.
In the AdS/CFT correspondence,
this solution can be understood as
a near horizon geometry of D3-branes
and three $U(1)$ charges are $R$-charges which
come from the Kaluza-Klein momenta in $S^5$.
Here, we consider the solution with all the three charges are equal.
We introduce perturbations around this solution,
and calculate the Green functions in the dual CFT by using 
the correspondence.
For simplicity, we impose a constraint such that
the three $U(1)$ gauge fields take the same value.
Then, the action of the STU model reduces to the following:
\begin{equation}
\label{action_bh}
S = \frac{1}{16\pi G_5} \int\!\dd^5x\sqrt{-g}
\Bigg( R + \frac{12}{l^2} \Bigg) +S_{\rm Maxwell} +S_{\rm CS} \ ,
\end{equation}
where the parameters $l$ is the radius of AdS space.
$S_{\rm Maxwell}$ is the action for $U(1)$ gauge field,
\begin{equation}
S_{\rm Maxwell}
= - \frac{1}{4e^2}\int\!\dd^5x \sqrt{-g} {\cal F}_{mn} {\cal F}^{mn} \ ,
\end{equation}
with ${\cal F}_{mn} = \partial_{m}{\cal A}_{n} - \partial_{n} {\cal A}_{m}$,
while $S_{\rm CS}$ is the Chern-Simons term,
\begin{equation}
S_{\rm CS}
= \frac{\kappa}{3} \int\!\dd^5x  \ \varepsilon^{lmnpq}
{\cal A}_l {\cal F}_{mn}{\cal F}_{pq} \ ,
\end{equation}
where $\kappa$ is the parameter,
through which the effect of the Chern-Simons term appears.

Almost same action without the Chern-Simons term can be
obtained from the bulk-filling D7-brane in the AdS background \cite{s}.
In the case of the bulk-filling D7-brane,
the $U(1)$ charge can be identified with the overall $U(1)$ part
of the $U(N_f)$ flavor symmetry, and
the gauge coupling
constant $e$ and the gravitational constant $G_5$ can be written as
\begin{equation}
\label{le2}
\frac{l}{e^2} = \frac{N_cN_f}{(2\pi)^2}  \qquad
{\rm and} \qquad  \frac{l^3}{G_5}= \frac{2 N_c^2}{\pi} \ ,
\end{equation}
where $N_c$ and $N_f$ denote numbers of D3- and D7-brane, respectively.
Even though the STU solution is not correspond to the SQCD which
comes from the D3-D7 system,
it can be expected that the dual field theory
shares some properties with the SQCD.

The equations of motion are given as
\begin{subequations}
\begin{eqnarray}
R_{mn}-\frac{1}{2}g_{mn}R  -\frac{6}{l^2}g_{mn}
&=&
8\pi G_5 T_{mn} \ ,
\label{eq_motion_E}
\\
-\frac{1}{e^2} \nabla_n {\cal F}^{mn}+\frac{\kappa}{\sqrt{-g}}
\varepsilon^{mlnpq} {\cal F}_{ln} {\cal F}_{pq}
&=&
0 \label{eq_motion_M} \ ,
\end{eqnarray}
\end{subequations}
where $T_{mn}$ is the energy-momentum tensor, 
\begin{equation}
T_{mn}=\frac{1}{e^2}\bigg({\cal F}_{mk}{\cal F}_{nl}g^{kl}
-\frac{1}{4}g_{mn}{\cal F}_{kl}{\cal F}^{kl}\bigg) \ .
\end{equation}
Since the presence of the Chern-Simons term does not change
the background spacetime from that in the bulk-filling
D7-brane model, it can be seen that the following
RN-AdS$_5$ background can be a classical solution,
\begin{eqnarray}
\label{rnads}
\begin{array}{rcl}
\dd s^2
&=&
\displaystyle \frac{r^2}{l^2} \Big( -f(r)(\dd t)^2 + 
(\dd\vec{x})^2\Big) + \frac{l^2}{r^2f(r)}(\dd r)^2 \ .
\\
{\cal A}_t
&=& \displaystyle - \frac{Q}{r^2} + \mu \ ,
\end{array}
\end{eqnarray}
with
\begin{eqnarray}
f(r) = 1-\frac{ml^2}{r^4}+\frac{q^2l^2}{r^6}
= \frac{1}{r^6}(r^2-r_0^2)(r^2-r_+^2)(r^2-r_-^2)
\ , \nonumber
\end{eqnarray}
and $q$ is related to $Q$ by
\begin{equation}
q
=
4\sqrt{\frac{\pi G_5}{3e^2}}~Q \ .
\end{equation}
$Q$ denotes the $U(1)$ charge induced from the bulk-filling D7-brane.
The parameters $m$ and $q$ correspond to the mass and charge of
the AdS space, respectively.
The explicit forms of $r_0(=-r_+^2-r_-^2)$ and $r_\pm$ are given by
\renewcommand{\arraystretch}{2.4}
\begin{eqnarray}
\begin{array}{rcl}
r_0^2
&=&
\displaystyle \left( \frac{m}{3q^2} \Bigg( 1+2\cos \bigg(
	       \frac{\theta}{3} +\frac{2}{3}\pi \bigg) \Bigg)
	      \right)^{-1} \ , \label{r0}
\\
r_+^2
&=&
\displaystyle \left( \frac{m}{3q^2} \Bigg(
	       1+2\cos\bigg(\frac{\theta}{3}+\frac{4}{3}\pi\bigg) \Bigg)
	      \right)^{-1}     \ , \label{r+}
\\
r^2_-
&=&
\displaystyle  \left( \frac{m}{3q^2} 
\Bigg( 1+2\cos\bigg(\frac{\theta}{3}\bigg) \Bigg) \right)^{-1}                   \ , \label{r-}
\end{array}
\end{eqnarray}
\renewcommand{\arraystretch}{1.7}
with
\begin{eqnarray}
\theta
=
\arctan \Bigg( \frac{3\sqrt{3}q^2\sqrt{\displaystyle
4m^3l^2-27q^4}}{2m^3l^2-27q^4} \Bigg) \ ,
\end{eqnarray}
where
$r_+$ and $r_-$ represent
locations of the outer and inner horizons, respectively.

Since the gauge potential ${\cal A}_t$ must vanish at the outer horizon,
the charge $Q$ and the chemical potential $\mu$ are related
as $Q = r_+^2 \mu$.
This background is induced by the back reaction of
the $U(1)$ charge as in \cite{cvetic2}.
In the case of the bulk-filling D7-brane,
the $U(1)$ charge is not identified as
the $U(1)$ baryon charge~\cite{s}.
Even though, there are no baryons in the field theory dual
of the STU solution,
we can expect that the baryon charge has similar property
to the $R$-charges in the STU solution.
In accordance with this analogy, $\mu$ which is the asymptotic value
of ${\cal A}_t$ can be interpreted as the chemical potential
for the $U(1)$ baryon charge.
Hence, we can interpret our results as a calculation for the QGP at
finite temperature in the presence of the $U(1)$ baryon charge.

Now there are two theories, which are obtained from the bulk-filling
D7-brane and the STU model as mentioned above.
The different point is the presence of the Chern-Simons term
in the later case.
The aim of this paper is to complete the linear order hydrodynamics
in the equations of motion for RN-AdS$_5$ with the Chern-Simons term.
The study without the Chern-Simons term has been
done~\cite{gmsst,gmssty}. 

For the well-defined variational principle in spacetime with 
boundary, 
we need the Gibbons-Hawking term defined as
\begin{eqnarray}
\label{gh}
S_{\rm GH}
&=&
\frac{1}{8\pi G_5} \! \int \! \dd^4x \sqrt{-g^{(4)}} K \ ,
\end{eqnarray}
where $g^{(4)}_{\mu\nu}$ and $K$ represent the four-dimensional 
induced metric and the extrinsic curvature on the boundary, 
respectively. 
The integration is taken on the boundary. 

The Hawking temperature of RN-AdS$_5$ background is given as
\begin{equation}
\label{temp}
T   =      \frac{r_+^2f'(r_+)}{4\pi l^2}
=      \frac{r_+}{\pi l^2}\bigg(1-\frac{1}{2}\frac{q^2l^2}{ r_+^6}\bigg)
\equiv \frac{1}{2\pi b}\Big(1-\frac{a}{2}\Big) \ ,
\end{equation}
where $a$ and $b$ are defined as
\begin{equation}
a \equiv \frac{q^2l^2}{r_+^6}
\qquad {\rm and} \qquad  b \equiv \frac{l^2}{2r_+} \ .
\end{equation}
We can see that the value of $a$ can be taken in $0 \le a < 2$.
These two parameters can be rewritten in terms of the chemical potential and
temperature as
\begin{equation} \label{ab}
a = 2 - \frac{4}{1+\sqrt{1+4(\tilde{\mu}/T)^2}}
\qquad {\rm and} \qquad 
b = \frac{1}{\pi T\big(1+\sqrt{1+4(\tilde{\mu}/T)^2}~\big)} \ ,
\end{equation}
where $\tilde{\mu}$ is a rescaled chemical potential
as $\tilde{\mu}\equiv\mu\sqrt{16\pi G_5/(3(\pi el)^2)}$
\footnote{For the R-charge,
$\tilde{\mu} \equiv \mu/ (2\sqrt{3}\pi)$, while for the brane charge,
$\tilde{\mu} \equiv \mu \sqrt{N_f/(3N_c\pi^2)}$.}.

\section{Hydrodynamics of RN-AdS$_5$ background}\label{sec:Perturb}

Let us now consider classical perturbations on the RN-AdS$_5$ background as
\begin{equation}
g_{mn} \equiv g^{(0)}_{mn}+h_{mn}
\quad {\rm and} \quad
{\cal A}_m \equiv A_m^{(0)}+A_m \ ,
\end{equation}
where $(g^{(0)}_{mn} \ {\rm and} \ A^{(0)}_m)$
and $(h_{mn}  \ {\rm and} \  A_m)$ denote
the RN-AdS$_5$ background (\ref{rnads}) and the perturbations, respectively.
We choose the following gauge conditions,
\begin{equation}
\label{GC}
h_{rm}=0 \quad \textrm{and} \quad A_r=0 \ ,
\end{equation}
and use Fourier expansion in which the momentum lies on the $z$-direction, 
\renewcommand{\arraystretch}{2.4}
\begin{eqnarray}
\begin{array}{rcl}
h_{\mu\nu}(t, z, r)
&=&
\displaystyle \!\int\!\frac{\dd^2k}{(2\pi)^2} \
\mbox{e}^{-i\omega t+ikz}h_{\mu\nu}(\omega,k, r) \ ,
\\
A_\mu(t, z, r)
&=&
\displaystyle \!\int\!\frac{\dd^2k}{(2\pi)^2} \
\mbox{e}^{-i\omega t+ikz} A_\mu(\omega,k, r) \ ,
\end{array}
\end{eqnarray}
where $\mu$ and $\nu$ run through the four-dimensional spacetime
except for the radial direction. 
The perturbations can be categorized  to the following three types,
\begin{itemize}
\item vector type
\ : \ $h_{xt}$,  \ $h_{yt}$,
\ $h_{xz}$,         \ $h_{yz}$
\ and \  $A_{x}$,  \ $A_{y}$
\item scalar type
\ : \ $h_{tt}$,  \ $h_{tz}$,
\ $h_{xx}=h_{yy}$,  \ $h_{zz}$
\ and \  $A_{t}$,  \ $A_{z}$
\item tensor type
\ : \ $h_{xy}$,  \ $h_{xx} = - h_{yy}$
\end{itemize}
Taking linear order for the perturbations in the equations of motion,
it turns out that the Chern-Simons term contributes to only the vector type.
Thus in this study, we consider the vector type.

\subsection{Decoupling of the equations of motion}

We start with introducing a dimensionless coordinate $u$
which is normalized by the outer horizon as
\begin{equation}
u \equiv \frac{r^2_+}{r^2} \ .
\end{equation}
In this notation, the horizon and the boundary are
located at $u=1$ and $0$, respectively.

Let us see the equations of motion (\ref{eq_motion_E})
and (\ref{eq_motion_M}) for the perturbation fields.
These are given as
\begin{subequations}
\begin{eqnarray}
0
&=&
{h^x_t}''  - \frac{1}{u}{h^x_t}' - \frac{b^2}{uf(u)}
\big( \omega k h^x_z + k^2h^x_t \big) - 3auB_x' \ ,
\label{eq_motion_v_001x} \\  							
0
&=&
{h^y_t}''  - \frac{1}{u}{h^y_t}' - \frac{b^2}{uf(u)}
\big( \omega k h^y_z + k^2h^y_t \big) - 3auB_y' \ ,
\label{eq_motion_v_001y} \\	
0
&=&
kf(u){h^x_z}' + \omega{h^x_t}'- 3a \omega u B_x \ ,
\label{eq_motion_v_002x} \\  		
0
&=&
kf(u){h^y_z}' + \omega{h^y_t}'- 3a \omega u B_y \ ,
\label{eq_motion_v_002y} \\
0
&=&
{h^x_z}'' + \frac{(u^{-1}f(u))'}{u^{-1}f(u)}{h^x_z}'
+ \frac{b^2}{uf^2(u)} \big( \omega^2h^x_z + \omega kh^x_t \big) \ ,
\label{eq_motion_v_003x} \\		
0
&=&
{h^y_z}'' + \frac{(u^{-1}f(u))'}{u^{-1}f(u)}{h^y_z}'
+ \frac{b^2}{uf^2(u)} \big( \omega^2h^y_z + \omega kh^y_t \big) \ ,
\label{eq_motion_v_003y}
\end{eqnarray}
and
\begin{eqnarray}
0
&=&
B_x'' + \frac{f'(u)}{f(u)}B_x' + \frac{b^2}{uf^2(u)}
\big( \omega^2 - k^2f(u) \big) B_x - \frac{1}{f(u)}{h^x_t}'
- \tilde{\kappa} \frac{i k}{f(u)}B_y \ ,
\label{eq_motion_v_004x} \\
0
&=&
B_y'' + \frac{f'(u)}{f(u)}B_y' + \frac{b^2}{uf^2(u)}
\big( \omega^2 - k^2f(u) \big) B_y - \frac{1}{f(u)}{h^y_t}'
+ \tilde{\kappa} \frac{i k}{f(u)}B_x \ ,
\label{eq_motion_v_004y}
\end{eqnarray}
\end{subequations}
with
\begin{eqnarray}
\label{fkb}
f(u)=(1-u)(1+u-au^2) \ ,
\quad
\tilde{\kappa} \equiv \frac{64 e^2 Q b^4}{l^5}\kappa \ ,
\quad
B_{x(y)} \equiv \frac{A_{x(y)}}{\mu}=\frac{l^4}{4Qb^2}A_{x(y)} \ , 
\quad
\end{eqnarray}
where the prime implies the derivative with respect to $u$.  
As we explain below, there are four independent variables.
(\ref{eq_motion_v_003x}) can be derived from
(\ref{eq_motion_v_001x}) and (\ref{eq_motion_v_002x}).
${h^x_z}'(u)$ could be expressed in terms of ${h^x_t}'(u)$ and $B_x(u)$
through (\ref{eq_motion_v_002x}).
From (\ref{eq_motion_v_001x}) and (\ref{eq_motion_v_002x})
we can obtain a second order differential equation for ${h^x_t}'(u)$ with
$B_x(u)$.
The same structure is hold for (\ref{eq_motion_v_001y}),
(\ref{eq_motion_v_002y}) and (\ref{eq_motion_v_003y}).
Together with (\ref{eq_motion_v_004x})
and (\ref{eq_motion_v_004y}),
we treat ${h^x_t}'(u)$, $B_x(u)$, ${h^y_t}'(u)$ and $B_y(u)$ as 
the four independent variables.

In order to solve the coupled equations of motion,
it might be convenient to introduce master variables~\cite{ki}.
We first take the following combinations~\cite{gmsst}:
\begin{eqnarray}
\label{master_variable_00}
\begin{array}{rcl}
\Theta_{x \pm}
&\equiv&
\displaystyle \frac{1}{u}{h^x_t}' - \Big( 3a - \frac{C_\pm}{u} \Big) B_x
\ ,
\\
\Theta_{y \pm}
&\equiv&
\displaystyle \frac{1}{u}{h^y_t}' - \Big( 3a - \frac{C_\pm}{u} \Big) B_y \ ,
\end{array}
\end{eqnarray}
with
$$ C_\pm \equiv 1+a \pm\sqrt{(1+a)^2+3ab^2k^2} \ .
$$
Using these variables,
we can obtain the following equations:
\begin{eqnarray}
\label{eq_motion_Phi_00p} \begin{array}{rcl}
0
&=&
\displaystyle  {\Theta}''_{x \pm} + 
\frac{(u^2f(u))'}{u^2f(u)}{\Theta}'_{x\pm}
+ \frac{b^2(\omega^2-k^2f(u))-uf(u)C_\pm}{uf^2(u)}{\Theta}_{x \pm}
- \frac{iC_\pm k \tilde{\kappa}}{C_0 f(u)} ({\Theta}_{y+} - {\Theta}_{y-})
\ ,
\\
0
&=&
\displaystyle  {\Theta}''_{y \pm} + 
\frac{(u^2f(u))'}{u^2f(u)}{\Theta}'_{y\pm}
+ \frac{b^2(\omega^2-k^2f(u))-uf(u)C_\pm}{uf^2(u)}{\Theta}_{y \pm}
+ \frac{iC_ \pm k \tilde{\kappa}}{C_0 f(u)} ({\Theta}_{x+} - {\Theta}_{x-}) \ ,
\end{array}
\end{eqnarray}
where $ C_0 \equiv C_+ - C_-$.
Comparing with the previous work~\cite{gmsst},
there still remain non-diagonal terms in the r.h.s of the equations
(\ref{eq_motion_Phi_00p}).
In order to diagonalize these equations,
we rewrite these by using a matrix form,
\begin{equation}
\label{eq_motion_Phi_01}
0
=
{\Theta}'' + \frac{(u^2f(u))'}{u^2f(u)} {\Theta}' + \Omega(u) {\Theta} \ ,
\qquad
{\rm with} \qquad
\Theta \equiv (\Theta_{x+},\Theta_{x-},\Theta_{y+},\Theta_{y-})^{\rm{T}}
\ ,
\end{equation}
and choose a matrix $\Lambda$ which leads a diagonal 
matrix $\widetilde{\Omega}\equiv\Lambda^{-1}\Omega\Lambda$ as  
\begin{equation}
\Lambda^{-1}
=\frac{1}{2C_0}
\left(
\begin{array}{cccc}
\ -i\frac{C_-\tilde{\kappa}}{D_+k^2} \ 
&
\ -i\frac{C_0^2-C_0D_+-2(1+a)\tilde{\kappa}k}{2D_+k^3} \ 
&
\ -\frac{C_-\tilde{\kappa}}{D_+k^2} \ 
&
\ -\frac{C_0^2-C_0D_+-2(1+a)\tilde{\kappa}k}{2D_+k^3} \ 
\\
i\frac{C_-\tilde{\kappa}k}{D_+}
&
i\frac{C_0^2+C_0D_+-2(1+a)\tilde{\kappa}k}{2D_+}
&
\frac{C_-\tilde{\kappa}k}{D_+}
&
\frac{C_0^2+C_0D_+-2(1+a)\tilde{\kappa}k}{2D_+}
\\
-i\frac{C_-\tilde{\kappa}}{D_-k^2}
&
i\frac{C_0^2-C_0D_-+2(1+a)\tilde{\kappa}k}{2D_-k^3}
&
\frac{C_-\tilde{\kappa}}{D_-k^2}
&
-\frac{C_0^2-C_0D_-+2(1+a)\tilde{\kappa}k}{2D_-k^3}
\\
i\frac{C_-\tilde{\kappa}k}{D_-}
&
-i\frac{C_0^2+C_0D_-+2(1+a)\tilde{\kappa}k}{2D_-}
&
-\frac{C_-\tilde{\kappa}k}{D_-}
&
\frac{C_0^2+C_0D_-+2(1+a)\tilde{\kappa}k}{2D_-}
\end{array}
\right),  
\end{equation}
with 
$$
D_{\pm} \equiv \sqrt{(C_0 \pm\tilde{\kappa} k)^2 \pm 4\tilde{\kappa}k{C_-}}
    =
2(1+a)\pm \tilde{\kappa} k +\frac{3ab^2}{1+a}k^2+{\cal O}(k^3)   \ .
$$
It might be important that 
the matrix $\Lambda$ is independent of $u$. 
Eigenvectors $\widetilde{\Theta} \equiv \Lambda^{-1} \Theta
\equiv(\widetilde{\Theta}_{1-}, \widetilde{\Theta}_{2-}, 
\widetilde{\Theta}_{1+}, \widetilde{\Theta}_{2+})^{\rm T}$ 
are not mixed by the derivatives with respect to $u$.
Moreover, defined new variables $\widetilde{\Theta}$ correspond 
to helicity bases on the $xy$-plane. 

The diagonalized equations are then given as
\begin{equation}
\label{eom}
0 = \widetilde{\Theta}''
+ \frac{(u^2f(u))'}{u^2f(u)} \widetilde{\Theta}'
+ \widetilde{\Omega}(u) \widetilde{\Theta}  \ ,
\end{equation}
with the diagonal matrix 
\begin{eqnarray}
\widetilde{\Omega}
&=&
\frac{b^2}{uf^2(u)}(\omega^2 - f(u)k^2) 
-\frac{1+a}{f(u)} 
\nonumber 
\\
&&+ \frac{1}{2f(u)}{\rm diag}
\Big(  - D_- + \tilde{\kappa} k , \  D_- + \tilde{\kappa} k , \
                                          - D_+ - \tilde{\kappa} k , \
					   D_+ - \tilde{\kappa} k
\Big). 
\label{tTheta}
\end{eqnarray}
This $\widetilde{\Theta}$ is a master variable with which we shall work.

\subsection{Solutions under the hydrodynamic approximation}\label{sec:sol}

The equation of motion (\ref{eom}) is an ordinary second order
differential equation with a regular singular point at $u=1$.
Thus, the solution can be generally written as
\begin{equation}
\label{general_sol}
\widetilde{\Theta}_i
=  (\widetilde{C}_1)_i (1-u)^{-\nu}  (\widetilde{\Phi}_1)_i
+ (\widetilde{C}_2)_i (1-u)^{ \nu} (\widetilde{\Phi}_2)_i
\quad \textrm{with} \quad \nu= i \frac{\omega}{4\pi T} \ ,
\end{equation}
where $\widetilde{C}_1$ and $\widetilde{C}_2$ are integral constants,
and $\widetilde{\Phi}_1(u)$ and $\widetilde{\Phi}_2(u)$ are regular
functions at $u=1$.
Index $i$ stands for the $i$-th component of vectors,
which takes $(1-)$, $(2-)$, $(1+)$ and $(2+)$.
The first term and the second term in the r.h.s. represent
a wave toward the horizon (in-going solution) and
a wave away from the horizon (out-going solution),
respectively\footnote{For a clear explanation, see ref.\cite{Son:2007vk}.}.
We choose the in-going solution ($\widetilde{C}_2 = 0$) as
the boundary condition at the horizon.
The remaining constant $\widetilde{C}_1$ will be fixed later.

We consider the expansion in which $k$ and $\omega$ are small comparing 
with the temperature.  
In this approximation, the perturbation variables describe the hydrodynamics
as the large distance and long time-scale classical wave on the equilibrium
state. 
In the vector mode, order of $k^2$ is the same as $\omega$.
In the present case, due to the Chern-Simons term, 
the expansion stars from $k$, which we call the first order.  
The second and the third orders are given by $(\omega, k^2)$ 
and $(k\omega, k^3)$, respectively. 
The expansion of $\widetilde{\Phi}_1(u)$ might be
\begin{eqnarray}
(\widetilde{\Phi}_1)_i  = \widetilde\rho_i (\widetilde X)_i
                        =  \widetilde\rho_i\left[(\widetilde{X}_{0})_i
                                + k        (\widetilde{X}_{k})_i
                                + \omega   (\widetilde{X}_{\omega})_i
                                + k^2      (\widetilde{X}_{k^2} )_i
                                + {\cal O}(k\omega, k^3)
                                \right] \ , \qquad
\label{expansion_Xp}
\end{eqnarray}
where we have introduced $\widetilde{\rho}_i(u)$ for later convenience.
Then, the equation of motion (\ref{eom}) for
$\widetilde{X}$ can be written in the following form,
\begin{equation}
\label{integral_Phip}
0= \bigg( u^2 f(u) \widetilde{\rho}_i{}^2(u) 
\Big((1-u)^{-\nu}(\widetilde{X})_i\Big)' \bigg)'
	+ u^2 f(u) \widetilde{\rho}_i(u) \ V_i(u) 
\ (1-u)^{-\nu}  (\widetilde{X})_i \ ,
\end{equation}
with
$$
V_i \equiv  \widetilde{\rho}_i\ {}''
+ \frac{(u^2f(u))'}{u^2f(u)}\widetilde{\rho}_i\ {}'
+ \widetilde{\Omega}_i(u) \widetilde{\rho}_i \ .
$$
We here take $\widetilde{\rho}_i(u)$ 
so that $V$ starts from the first order in the expansion 
of $k$ and $\omega$,
$$
\widetilde{\rho} \equiv
\Big(-\frac{2(1+a)}{u}+3a,\ 1,\ -\frac{2(1+a)}{u}+3a,\ 1 \Big)^{\rm T} \ .
$$
Then we could solve the differential equation order by order.
The equation of each order takes the form of
$(F(\widetilde X_*)_i')' = G$ schematically. 
Here, $F$ is a function of $u$, and $G$ contains only
lower orders of $\widetilde X_*$.
Therefore 
we can obtain the solution by integrating these equations twice.

Two integration constants in this procedure
are related to the two integration constants in (\ref{general_sol}).
We have imposed an in-going condition at the horizon and
the singularity has been already factorized.
Then, one of the integration constants in $\widetilde X_*(u)$ 
must be taken such that they are regular.
The other integration constant can be absorbed
into the normalization constant $\widetilde C_1$. 
In order to fix the constant $\widetilde{C}_1$, 
we use the following relation given by 
(\ref{eq_motion_v_001x}), (\ref{eq_motion_v_001y})
and (\ref{master_variable_00}),
\renewcommand{\arraystretch}{2.4}
\begin{eqnarray}
\begin{array}{rcl}
\label{babc}
\displaystyle u^2 \Theta_{x\pm}' - u C_\pm B_x' \Big|_{u=0}
\!\!\!
&=&
b^2\Big(\omega k (h^x_z)^0+ k^2 (h^x_t)^0\Big) - C_\pm(B_x)^0 \ ,
\\
\displaystyle u^2 \Theta_{y\pm}' - u C_\pm B_y' \Big|_{u=0}
\!\!\!
&=&
b^2\Big(\omega k (h^y_z)^0+ k^2 (h^y_t)^0\Big) -C_\pm(B_y)^0 \ , 
\end{array}
\end{eqnarray}
where $(h^{x(y)}_t)^0$, $(h^{x(y)}_z)^0$ and $(B_{x(y)})^0$ mean 
constant values at the boundary.
We rewrite the l.h.s. of these equations in terms 
of $\widetilde \Phi_1(u)$ and substitute the solution of
the differential equations. 
Solving the equations (\ref{babc}), 
we can then obtain the expression of $\widetilde{C}_1$
in terms of the boundary values of perturbations variables.

Following the procedure above, we can obtain the solutions 
around the boundary up to ${\cal O}(\omega, k^2)$: 
\begin{subequations}
\begin{eqnarray}
\label{sol:hxit}
h^x_t (u)
&=&
(h^x_t)^0 - b^2\Big(k^2 (h^x_t)^0+\omega k (h^x_z)^0\Big)u 
\nonumber 
\\
&&
+\frac{1}{8P}
\Bigg\{
4b\Big(k^2(h^x_t)^0+\omega k(h^x_z)^0\Big)
+3ia
\Big(
4\omega
-\frac{3ab^2}{(1+a)^2}\omega k^2
\Big)(B_x)^0
\Bigg\}u^2 \ , 
\\
\label{sol:hyit}
h^y_t (u)
&=&
(h^y_t)^0 - b^2\Big(k^2 (h^y_t)^0+\omega k (h^y_z)^0\Big)u 
\nonumber 
\\
&&
+\frac{1}{8P}
\Bigg\{
4b\Big(k^2(h^y_t)^0+\omega k(h^y_z)^0\Big)
+3ia
\Big(
4\omega
-\frac{3ab^2}{(1+a)^2}\omega k^2
\Big)(B_y)^0
\Bigg\}u^2 \ , 
\\
\label{sol:hxiz}
h^x_z(u)
&=&
(h^x_z)^0 
+b^2\Big(\omega^2(h^x_z)^0+\omega k(h^x_t)^0\Big)u
\nonumber 
\\
&&
-\frac{1}{4P}
\Bigg\{
2b\Big(\omega^2(h^x_z)^0+\omega k(h^x_t)^0\Big)
+\frac{3ab}{1+a}\omega k (B_x)^0
\Bigg\}u^2 \ , 
\\
\label{sol:hyiz}
h^y_z(u)
&=&
(h^y_z)^0 
+b^2\Big(\omega^2(h^y_z)^0+\omega k(h^y_t)^0\Big)u
\nonumber 
\\
&&
-\frac{1}{4P}
\Bigg\{
2b\Big(\omega^2(h^y_z)^0+\omega k(h^y_t)^0\Big)
+\frac{3ab}{1+a}\omega k (B_y)^0
\Bigg\}u^2 \ , 
\\
\label{sol:Bx}
B_x(u) 
&=&
(B_x)^0 +b^2k^2 (B_x)^0 u\log u
\nonumber 
\\
&&
+\frac{i}{8(1+a)^2P}
\Bigg\{
\frac{1}{1+a}
\Big(
-12a(1+a)^2\omega
+2i(1+a)(2-a)b\omega^2
\nonumber 
\\
&&
\hspace*{40.5mm}
-(12+8a-17a^2-4a^3)b^2\omega k^2
\Big)(B_x)^0
\nonumber 
\\
&&
\hspace*{26mm}
+\frac{\kappa}{2(1+a)}
\Big(
2(1+a)(4+a)bk^3
-4i(4+2a+a^2)\omega k
\Big)
(B_y)^0
\nonumber 
\\
&&
\hspace*{26mm}
+
\Big(
4i(1+a)bk^2
+6ab^2\omega k^2
\Big)(h^x_t)^0
+4i(1+a)b\omega k (h^x_z)^0
\nonumber 
\\
&&
\hspace*{26mm}
+2\kappa b\Big(k^3(h^y_t)^0
+\omega k^2(h^y_z)^0\Big)
\Bigg\}u \ , 
\\
\label{sol:By}
B_y(u) 
&=&
(B_y)^0 +b^2k^2 (B_y)^0 u\log u
\nonumber 
\\
&&
+\frac{i}{8(1+a)^2P}
\Bigg\{
\frac{1}{1+a}
\Big(
-12a(1+a)^2\omega
+2i(1+a)(2-a)b\omega^2
\nonumber 
\\
&&
\hspace*{40.5mm}
-(12+8a-17a^2-4a^3)b^2\omega k^2
\Big)(B_y)^0
\nonumber 
\\
&&
\hspace*{26mm}
-\frac{\kappa}{2(1+a)}
\Big(
2(1+a)(4+a)bk^3
-4i(4+2a+a^2)\omega k
\Big)
(B_x)^0
\nonumber 
\\
&&
\hspace*{26mm}
+
\Big(
4i(1+a)bk^2
+6ab^2\omega k^2
\Big)(h^y_t)^0
+4i(1+a)b\omega k (h^y_z)^0
\nonumber 
\\
&&
\hspace*{26mm}
-2\kappa b\Big(k^3(h^x_t)^0
+\omega k^2(h^x_z)^0\Big)
\Bigg\}u \ , 
\end{eqnarray}
\end{subequations}

\vspace*{-6mm}
\noindent
where $P$ gives a pole structure with a diffusion constant $D$,  
\begin{equation}
P=i\omega-Dk^2=i\omega-\frac{b}{2(1+a)}k^2. 
\label{pole_0}
\end{equation}
We can observe the effect of Chern-Simons term in $B_{x(y)}(u)$ via
the parameter $\kappa$. 

It should be worth to mention the different structure in the 
equation of motion (\ref{eom}), which may be crucial to obtain 
analytical solutions. 
The first and the third components 
of (\ref{eom}) can be solved analytically up to the second order 
${\cal O}(\omega, k^2)$, while 
the second and the fourth components 
can be solved up to the third order ${\cal O}(k\omega, k^3)$. 
Since the second and the fourth components provide the pole structure, 
the equation (\ref{pole_0}) would be modified to 
\begin{equation}
\label{P}
{\cal P} = 
i \omega - D k^2
\pm \frac{ab}{4(1+a)^2} \tilde{\kappa}k^3 \ , 
\end{equation}
where $\pm$ correspond to the helicity. 
We can observe that the Chern-Simons term now affects to the pole 
structure.
This was firstly calculated in~\cite{Sahoo:2009yq}.  
Our result is consistent with theirs.  

\section{The dual quark-gluon-plasma}\label{sec:dGQP}

We have been studying the dual gravity side so far 
and obtained the solutions.
In this section, we move to the field theory side and
consider the effect of the Chern-Simons term to the retarded
two-point Green function in the dual QGP at finite
temperature with the $U(1)$ baryon density.

\subsection{Boundary action}

In order to obtain the retarded two-point Green function 
following the GKP-W relation~\cite{gkp,w},
we need the bilinear-part of the regularized boundary action at $u=0$.
Together with the Gibbons-Hawking term (\ref{gh}),  
we need the following counter term for the regularized
action at the boundary, 
\begin{eqnarray}
\label{ct}
S_{\rm ct} = S_{\rm ct, \ gravity}+S_{\rm ct, \ gauge} \ ,
\end{eqnarray}
where 
\begin{subequations}
\begin{eqnarray}
S_{\rm ct,\ gravity} 
&=&
\frac{1}{8\pi G_5} \! \int \! \dd^4x \sqrt{-g^{(4)}} 
\bigg(\  \frac{3}{l} - \frac{l}{4} R^{(4)} \bigg) \ , 
\\
S_{\rm ct,\ gauge}   
&=& 
 \frac{l}{8e^2} \log u \! \int \! \dd^4x \sqrt{-g^{(4)}} 
\ {\cal F}_{mn} {\cal F}^{mn} \ ,
\end{eqnarray}
\end{subequations}

\vspace*{-6mm}
\noindent
where $R^{(4)}$ is the curvature on the boundary. 
$S_{\rm ct,\ gravity}$ is given in \cite{bk}.
On the other hand, $S_{\rm ct,\ gauge}$ is obtained 
to cancel the logarithmic divergence 
coming from gauge field fluctuations. 

The boundary action derived from (\ref{action_bh}) is
\begin{eqnarray}
\label{sgra}
S^{(0)} =
\frac{l^3}{256\pi b^4 G_5 u^2} \!\int\!\frac{\dd^2k}{(2\pi)^2}
\Bigg\{
&&
\frac{u f'(u)}{f(u)}   (h^x_t)^2
+   h^x_t        \Big(h^x_t - 3 u {h^x_t}'\Big)
- f(u) h^x_z     \Big(h^x_z - 3 u {h^x_z}'\Big)
\nonumber \\
&&
+ 3a B_x \Big(h^x_t - f(u) {B_x}'\Big) \Bigg\} \ .
\end{eqnarray}
The Gibbons-Hawking term (\ref{gh}) and the counter term (\ref{ct}) are 
\begin{eqnarray}
\label{sgh}
S^{(0)}_{\rm GH}
&=&
\frac{l^3}{256\pi b^4G_5 u^2} \!\int\!\frac{\dd^2k}{(2\pi)^2}
\Bigg\{ - \frac{u f'(u)}{f(u)} (h^x_t)^2 - 4 (h^x_t)^2
- u f'(u) (h^x_z)^2 + 4 u h^x_t {h^x_t}'  \nonumber \\
&& \hspace*{38mm}
+ 4 f(u) \Big( (h^x_z)^2 - u h^x_z(h^x_z)' \Big)  \Bigg\} \ ,   \\
\label{sct}
S^{(0)}_{\rm ct}
&=&
\frac{3 l^3}{256\pi b^4G_5 u^2\sqrt{f(u)}} \!\int\!\frac{\dd^2k}{(2\pi)^2}
\Big\{ (h^x_t)^2 - f(u) (h^x_z)^2 
\nonumber 
\\
&&
\hspace*{50mm}
+ab^2k^2 u^2f(u)\log u \Big(B_x^2+B_y^2\Big)
\Big\} \ .
\end{eqnarray}
Then, the regularized boundary action is given as:
\begin{eqnarray}
\label{surface}
S_{\rm st}
&=&
\lim_{u \to 0}(S^{(0)} + S^{(0)}_{\rm GH} + S^{(0)}_{\rm ct} ) \nonumber
\\
&=& \lim_{u \to 0} \frac{l^3}{256\pi b^4 G_5} \!\int\!\frac{\dd^2k}{(2\pi)^2}
\Bigg\{
\frac{1}{u}\Big(h^x_t(-k){h^x_t}'(k)-h^x_z(-k){h^x_z}'(k)\Big)
\nonumber
\\
&&
\hspace*{40mm}
+ \frac{1+a}{2}\Big( 3h^x_t(-k)h^x_t(k)+ah^x_z(-k)h^x_z(k) \Big)
\nonumber \\ &&
\hspace*{40mm}
- 3a B_x(-k)\Big( h^x_t(k)+{B_x}'(k)\Big)
\nonumber 
\\
&&
\hspace*{40mm}
+3ab^2k^2
\log u \Big(B_x(-k)B_x(k)+B_y(-k)B_y(k)\Big)
\Bigg\} \ .
\end{eqnarray}
We can see that the Chern-Simons term does not explicitly enter
the boundary action.
This can be expected from the original property of the Chern-Simons term.

\subsection{Retarded two-point Green functions}

Substituting the solutions (\ref{sol:hxit})-(\ref{sol:By}) to the
regularized boundary action (\ref{surface}),
we can obtain the retarded two-point Green functions 
up to ${\cal O}(\omega, k^2)$ as
\begin{subequations}
\begin{eqnarray}
G_{xt \ xt}(k,\omega) 
= 
G_{yt \ yt}(k,\omega) 
&=&
\frac{l^3}{128 \pi b^3 G_5} 
\frac{k^2}{P} \ , 
\label{Gxtxt} 
\\
G_{xt \ xz}(k, \omega)
=G_{yt \ yz}(k, \omega)
&=&
-\frac{l^3}{128\pi b^3G_5}\frac{\omega k}{P} \ , 
\label{Gxtxz}
\\
G_{xz \ xz}(k, \omega)
=G_{yz \ yz}(k, \omega)
&=&
\frac{l^3}{128 \pi b^3 G_5} 
\frac{\omega^2}{P} \ , 
\label{Gxzxz}
\\
G_{xt \ x}(k,\omega) 
= G_{yt \ y}(k,\omega)
&=&
-\frac{Q}{4 e^2 l^3 (1+a)}\Bigg(\frac{2bk^2 +4i(1+a)\omega}{P}\Bigg)
\ , 
\label{Gxtx} 
\\
G_{xz \ x}(k, \omega)
=G_{yz \ y}(k, \omega)
&=&
\frac{Qb}{e^2l^3(1+a)}
\frac{\omega k}{P} \ ,
\label{Gxzx}
\\
G_{x \ yt}(k, \omega)
=-G_{y \ xt}(k, \omega)
&=&
-\frac{768Q^2b^5}{(1+a)l^8}\frac{k^3}{P}\kappa
\ , 
\label{Gyxt}
\\
G_{x \ x}(k,\omega) 
= G_{y \ y}(k,\omega)
&=&
\frac{3ial}{4e^2(1+a)b^2}\frac{\omega}{P}
\ , 
\label{Gxx}
\\
G_{x \ y}(k,\omega)   
= -G_{y \ x}(k,\omega)
&=&
\frac{6Qb^2}{(1+a)^2l^4}
\left(\frac{i(4+a)bk^3+2(4+2a+a^2)\omega k}{P}\right)\kappa  
\ , \ \  
\label{Gyx}
\\
{\mbox{(others)}}
&=&
0 \ , 
\end{eqnarray}
\end{subequations}

\vspace*{-6mm}
\noindent
where $P$ is given by (\ref{pole_0}), which gives the pole structure.
In the above expressions, we restored the gauge field $A_{x(y)}$ 
and $\kappa$ by using the relations (\ref{fkb}) 
and rise and lower the indices by using the boundary Minkowski metric.
It can be seen that the Chern-Simons term 
affects $G_{x \ yt}(k, \omega)$ and $G_{x \ y}(k, \omega)$ 
via $\kappa$.

The presence of non-vanishing $G_{x \ y}(k, \omega)$ 
indicates 
the existence of chiral magnetic conductivity 
introduced in~\cite{kharzeev}, which 
is defined by $J_i=\sigma_{B} {\cal B}_i$ where ${\cal B}_i$ is 
the magnetic field.   
Using ${\cal B}_x(k)=(-ik) A_y(k)$,   
$\sigma_B$ in hydrodynamic limit can be read off: 
\begin{eqnarray}
\sigma_{B}(\omega, k)
&=&
\frac{1}{-ik} G_{x\ y}(k, \omega)
\nonumber 
\\ 
&=&
-\frac{6Qb^2}{(1+a)^2l^4P}
\Big(-(4+a)bk^2+2i(4+2a+a^2)\omega\Big)\kappa \ .  
\label{mc}
\end{eqnarray}
Taking the limit $\omega\to0$ first and then $k\to 0$, 
the DC conductivity $\sigma_B^0$  becomes 
\begin{equation}
\sigma_B^0
=\lim_{\omega\to0}
\sigma_B(\omega, k)
=-\frac{12Q(4+a)b^2}{l^4(1+a)}\kappa \ .
\label{mc0}
\end{equation}
The presence of magnetic conductivity is the most important 
consequence of the Chern-Simons term.
The physical origin is due to the effect of the $U(1)_R$ anomaly. 
In \cite{kharzeev} the magnetic conductivity due to the 
axial anomaly  was discussed: 
The magnetic field can change the spin direction and the momentum is 
tied with the spin, making left and right handed 
particles move in opposite direction.  
Therefore the magnetic field can induce an electric current 
proportional to the difference of left and right handed zero modes, 
which is nothing but the chiral anomaly.  
In our case the role of chiral symmetry is replaced 
by the $U(1)_R$ symmetry. 

The magnetic conductivity in holographic set-up was 
calculated in \cite{Son:2009tf, Yee:2009vw}. 
Our result agrees with \cite{Son:2009tf}
\footnote{
In terms of the variables in \cite{Son:2009tf},  
our magnetic conductivity (\ref{mc0}) should be compared with  $\xi_B$
and two are related by $\sigma_B^0= \frac 3 2 \xi_B$, where   
$\xi_B =  
-\frac{\sqrt{3}(3R^4+m)q\kappa} {4\pi G_5mR^2}$, 
with $R$ being the horizon radius.  
} 
but not with \cite{Yee:2009vw}. 
The difference can be traced to that of the set-ups\footnote{ 
In \cite{Yee:2009vw}, the author introduced two gauge fields 
which are associated with the axial and vector $U(1)$.  
He considers non-zero background charge 
for the axial $U(1)$ gauge field 
and perturbations only for the gauge field of vector $U(1)$.  
Then, the perturbation of gravity is 
decoupled from that of the gauge field. 
On the other hand, in our setup, 
we considered the background charge and 
perturbations for the same gauge field, 
and then, the gravitational perturbation 
couples to the gauge field perturbation.
As a consequence, the DC magnetic conductivity of \cite{Yee:2009vw}
 is independent of temperature while our result depends on it. }.     

It is interesting to express the magnetic conductivity in 
terms of the boundary variables, 
i.e.  the temperature and the chemical potential through the equations 
(\ref{ab}):   
\begin{eqnarray}
 \sigma_{B}^0  = -  6\kappa \mu  (\widetilde{\sigma}_{B})_0 , \qquad        
    (\widetilde{\sigma}_{B})_0 = 
\frac{ 3 \gamma +1 } 
{  3\gamma -1 }, 
\label{mc2}
\end{eqnarray}
where $\gamma=\sqrt{1+4(\tilde{\mu}/T)^2}$. 
The  behaviors in low 
and high temperature regions are given by 
\begin{subequations}
\begin{eqnarray}
 (\widetilde{\sigma}_{B})_0 &=& 
              1
+  \frac{1}{3}   \frac{T}{\tilde{\mu}}    
+  \frac{1}{18} \left( \frac{T}{\tilde{\mu}} \right)^2 
+   {\cal O}\left( (T/\tilde{\mu})^3 \right)   
\quad \textrm{: low $T$} \ , \quad\qquad \\
 (\widetilde{\sigma}_{B})_0 &=& 
   2 - 3\left( \frac{\tilde{\mu}}{T} \right)^2 
+ {\cal O}\left( (\tilde{\mu}/T)^6 \right)  
\quad \textrm{: high $T$}  \ . \quad  
\end{eqnarray}
\end{subequations} 

\vspace*{-6mm}

It should be noticed that at low temperature,
 the magnetic conductivity   increases  
linearly in $T$ starting from a non-vanishing value and 
saturate to a finite value. 
Interestingly, its value at the large temperature is just 
twice of the value at zero temperature.  
Such behavior should be contrasted with that of  electric conductivity
 \cite{gmssty} 
\begin{eqnarray}
\sigma_E = \frac{l e_E^2}{8e^2} 
\frac{(2-a)^2}{b(1+a)^2}\equiv \frac{l e_E^2}{8e^2}\cdot 8\pi {\tilde \mu} 
(\widetilde{\sigma}_{E})_0, 
\end{eqnarray} 
 which is $\sim T^2$ and $\sim T$ in low and high temperature 
 regime respectively, as one can see from 
\begin{subequations}
\begin{eqnarray}
 (\widetilde{\sigma}_{E})_0 &=& 
   \frac{1}{9} \left(\frac{T}{\tilde{\mu}}\right)^2
+\frac{5}{54} \left(\frac{T}{\tilde{\mu}}\right)^3
+   {\cal O}\left( (T/\tilde{\mu})^5 \right)   
\quad \textrm{: low $T$} \ , \quad\qquad \\
 (\widetilde{\sigma}_{E})_0 &=& 
  \frac{T}{\tilde{\mu}}  - {5}{\frac{\tilde{\mu}}{T}}
+26 \left( {\frac{\tilde{\mu}}{T}}\right)^3    
 + {\cal O}\left( (\tilde{\mu}/T)^5 \right)  
\quad \textrm{: high $T$}  \ . \quad  
\end{eqnarray}
\end{subequations} 

\begin{figure}[htbp]
\begin{center}
\includegraphics*[scale=0.6]{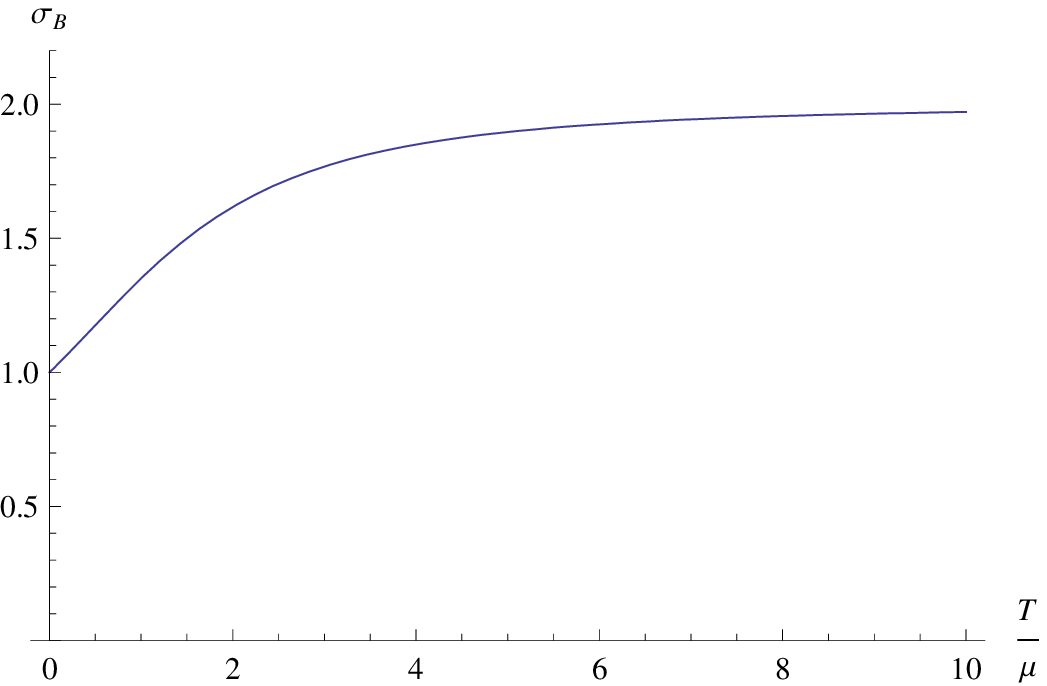}
\includegraphics*[scale=0.6]{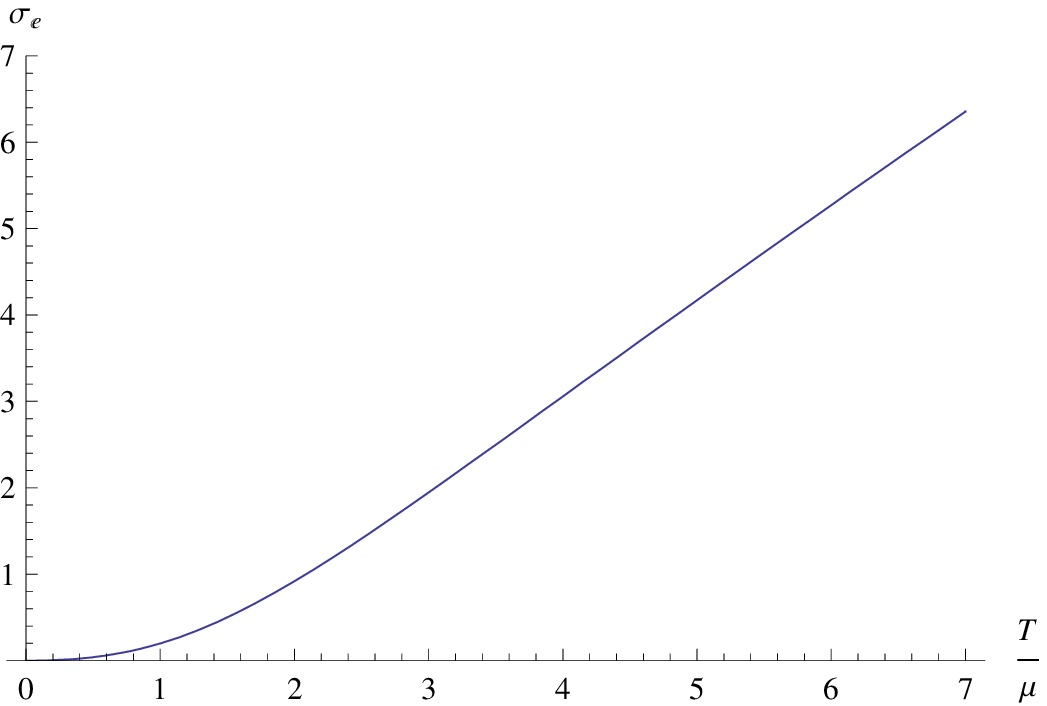}
\end{center}
\caption{ conductivities as functions of  $T/{\tilde\mu}$. 
Left figure is $(\widetilde{\sigma}_B)_0$ and   
right one is $(\widetilde{\sigma}_E)_0$.   }
\label{fig}
\end{figure}

\vspace*{-6mm}
\noindent
Whole behavior of magnetic and electric conductivities are 
drawn as functions of the $T/\tilde\mu$ 
in Figure~\ref{fig}.  

\section{Summary}\label{sec:summary}
 The RN-AdS$_5$ background may be considered as is the dual
gravity for the QGP with finite temperature and the baryon density.
One can add the Chern Simon  term to the action.
Although it does not change the solution, it can affect the 
fluctuation around the RN black hole. 
We have studied its effect in the hydrodynamics of the QGP coming.
Although this interpretation is interesting from the perspective 
of more realistic holographic QCD,
the theory is better known to come from the 10 dimensional STU 
solution with diagonal U(1) charges.  
From this point of view,   
the Chern-Simons term has a predetermined coupling. 

We worked out the decoupling problem of 
equations of motion by using
the master variables in the hydrodynamic analysis on the dual gravity.
As a result,   one of the effects of 
the Chern-Simons term 
appears in the retarded two-point Green function of $U(1)$
currents $G_{y \ x}(k, \omega)$ and $G_{x \ y}(k, \omega)$ in 
the dual QCD, which  lead to the magnetic conductivity.  
This quantity is written in terms of the temperature and the chemical 
potential and compared with electric conductivity. 
The Chern-Simons term 
also contributes to the hydrodynamic pole structure in ${\cal O}(k^3)$.  

In the case of the D7-brane effective theory, 
the Chern-Simons term in the five-dimensional gauge theory appears 
if we introduce the 3-form RR-flux which lies on the $S^3$ part 
of the world volume.
The 3-form flux is the magnetic flux associated with the D5-brane, 
and baryons can be constructed by using the 
D5-brane~\cite{Witten:1998xy}.
It would be interesting to consider the relation between 
the effect of the Chern-Simons term and the magnetic flux of 
the D5-brane. 

\vspace*{5mm}

\noindent{\large{\bf Acknowledgments}}

\vspace*{2mm}
\noindent
We would like to thank  Kwang-Hyun Jo, Masaki Murata, Tatsuhiro Misumi,
Shin Nakamura, Ho-Ung Yee and members of Friday Meeting in CQUeST.
The work of SJS was supported in part by KOSEF Grant
R01-2007-000-10214-0 and SRC Program of the NRF
through the CQUeST with grant number 2005-0049409.
TT acknowledges the Max Planck Society (MPG), the Korea Ministry
of Education, Science and Technology (MEST) and
POSTECH for the support of the Independent Junior Research Group
at APCTP.

\appendix

\section{Retarded Green function in AdS/CFT}
\label{app:Green}

In this appendix, we briefly summarize the retarded Green
function of the boundary theory in AdS/CFT correspondence
formulated by Son and Starinets~\cite{ss}.
To begin with, we consider perturbations on five-dimensional background.
We refer $x^\mu$ and $u$ as the four-dimensional and the radial
coordinates, respectively,
The boundary and the horizon locate at $u=0$ and $u=1$, respectively.
We now suppose a solution of the equation of motion can be written as
\begin{equation}
\phi(u,x) = \!\int\!\frac{\dd^4 k}{(2\pi)^4}\ \mbox{e}^{ikx}f_k(u) {\phi}(k)
\end{equation}
A quadratic part of the surface action can be written as
\begin{equation}
\label{on_shell_action}
S[\phi]
= \!\int\!\frac{\dd^4 k}{(2\pi)^4} \phi(-k)G(k, u)\phi(k)\bigg|_{u=0}^{u=1} \ ,
\end{equation}
where $G(k,u)$ is a function of $f_{\pm k}(u)$
and $\partial_u f_{\pm k}(u)$.
Applying GKP-W relation~\cite{gkp,w}
to spacetimes in the real-time,
the formula for the two-point retarded Green functions
has been defined as
\begin{equation}
\label{Green_function}
G^{\rm R}(k) \equiv 2G(k, u) \bigg|_{u=0} \ .
\end{equation}

Lastly, we define the retarded Green functions we discuss in this paper:
\renewcommand{\arraystretch}{2.4}
\begin{eqnarray}
\label{diff_Green_function}
\begin{array}{rcl}
G_{\mu\nu \ \rho\sigma}(k,\omega) &\equiv& \displaystyle
-i\!\int\!\frac{\dd^2x}{(2\pi)^2} \ \mbox{e}^{-i\omega t +i kz}\theta(t) \
\Big\langle {[} T_{\mu\nu}(t,z), \ T_{\rho\sigma}(0,0) {]} \Big\rangle \ , \\
G_{\mu\nu \ \rho}(k,\omega) &\equiv& \displaystyle
-i\!\int\!\frac{\dd^2x}{(2\pi)^2} \ \mbox{e}^{-i\omega t +i kz}\theta(t) \
\Big\langle{[} T_{\mu\nu}(t,z),  \ J_\rho(0,0)         {]} \Big\rangle \ , \\
G_{\mu \ \nu}(k,\omega) &\equiv& \displaystyle
-i\!\int\!\frac{\dd^2x}{(2\pi)^2} \ \mbox{e}^{-i\omega t +i kz}\theta(t) \
\Big\langle{[} J_\mu(t,z),       \ J_\nu(0,0)          {]} \Big\rangle \ ,
\end{array}
\end{eqnarray}
where the momentum in this paper is taken to $z$-direction,
and the operators $T_{\mu\nu}(t,z)$ and $J_\mu(t,z)$ are the
energy-momentum tensor and the $U(1)$ baryon current, respectively.



\begin{thebibliography}{99}
\bibitem{ads/cft}
J.M. Maldacena,
Adv. Theor. Math. Phys. {\bf 2} (1998) 231,
{\tt [arXiv:hep-th/9711200]}.

\bibitem{gkp}
S.S. Gubser, I.R. Klebanov and A.M. Polyakov,
Phys.\ Lett.\ {\bf B428} (1998) 105,
{\tt [arXiv:hep-th/9802109]}.

\bibitem{w}
E. Witten,
Adv.\ Theor.\ Math.\ Phys.\ {\bf 2} (1998) 253,
{\tt [arXiv:hep-th/9802150]}.

\bibitem{pss0}
G. Policastro, D.T. Son and A.O. Starinets,
Phys.\ Rev.\ Lett.\  {\bf 87} (2001) 081601,
{\tt [arXiv:hep-th/0104066]}.

\bibitem{ksz}
K.-Y. Kim, S.-J. Sin and I. Zahed,
{\tt [arXiv:hep-th/0608046]}.

\bibitem{ht}
N. Horigome and Y. Tanii,
JHEP {\bf 0701} (2007) 072,
{\tt [arXiv:hep-th/0608198]}.

\bibitem{nssy1}
S. Nakamura, Y. Seo, S.-J. Sin and K.P. Yogendran,
J. Korean Phys. Soc. {\bf 52} (2008) 1734, 
{\tt [arXiv:hep-th/0611021]}.

\bibitem{kmmmt}
S. Kobayashi, D. Mateos, S. Matsuura, R.C. Myers and R.M. Thomson,
JHEP {\bf 0702} (2007) 016,
{\tt [arXiv:hep-th/0611099]}.

\bibitem{nssy2}
S. Nakamura, Y. Seo, S.-J. Sin and K.P. Yogendran,
Prog. Theor. Phys. {\bf 120} (2008) 51, 
{\tt [arXiv:0708.2818[hep-th]]}.

\bibitem{bergman}
O. Bergman, G. Lifschytz and M. Lippert,
JHEP {\bf 0711} (2007) 056,
{\tt [arXiv:0708.0326 [hep-th]]}.

\bibitem{ubc}
M. Rozali, H.H. Shieh, M. Van Raamsdonk and J. Wu,
JHEP {\bf 0801} (2008) 053,
{\tt [arXiv:0708.1322[hep-th]]}.

\bibitem{n}
S. Nakamura, Prog. Theor. Phys. {\bf 119} (2008) 839,
{\tt [arXiv:0711/1601[hep-th]]}.

\bibitem{ssz}
E. Shuryak, S.-J. Sin and I. Zahed,
J.\ Korean Phys.\ Soc.\  {\bf 50} (2007) 384,
\\
{\tt [arXiv:hep-th/0511199]}.

\bibitem{ss}
D.T. Son and A.O. Starinets,
JHEP {\bf 0209} (2002) 042, 
{\tt [arXiv:hep-th/0205051]}.

\bibitem{pss}
G. Policastro, D.T. Son and A.O. Starinets,
JHEP {\bf 0209} (2002) 043, \\
{\tt [arXiv:hep-th/0205052]}.

\bibitem{pss2}
G. Policastro, D.T. Son and A.O. Starinets,
JHEP {\bf 0212} (2002) 054, \\
{\tt [arXiv:hep-th/0210220]}.

\bibitem{hs}
C.P. Herzog and D.T. Son,
JHEP {\bf 0303} (2003) 046, 
{\tt [arXiv:hep-th/0212072]}.

\bibitem{ks}
P.K. Kovtun and A.O. Starinets,
Phys. Rev. {\bf D72} (2005) 086009, \\
{\tt [arXiv:hep-th/0506184]}.

\bibitem{gmsst}
X.-H. Ge, Y. Matsuo, F.-W. Shu,
S.-J. Sin and T. Tsukioka,
Prog. Theor. Phys. {\bf 120} (2008) 833,
{\tt[arXiv:0806.4460[hep-th]]}.

\bibitem{gmssty}
Y. Matsuo, S.-J. Sin, S. Takeuchi, T. Tsukioka and C.-M. Yoo,
Nucl.\ Phys.\ {\bf B820} (2009) 593,
{\tt [arXiv:0901.0610[hep-th]]}.

\bibitem{gmsst2}
X.-H. Ge, Y. Matsuo, F.-W. Shu, S.-J. Sin and T. Tsukioka,
JHEP {\bf 0810} (2008) 009, 
{\tt[arXiv:0808.2354[hep-th]]}.

\bibitem{s}
S.-J. Sin,
JHEP {\bf 0710} (2007) 078, 
{\tt [arXiv:0707.2719[hep-th]]}.

\bibitem{mas}
J. Mas,
JHEP {\bf 0603} (2006) 016, 
{\tt [arXiv:hep-th/0601144]}.

\bibitem{ss2}
D.T. Son and A.O. Starinets,
JHEP {\bf 0603} (2006) 052, 
{\tt [arXiv:hep-th/0601157]}.

\bibitem{mno}
K. Maeda, M. Natsuume and T. Okamura,
Phys. Rev. {\bf D73} (2006) 066013, \\
{\tt [arXiv:hep-th/0602010]}.

\bibitem{saremi}
O. Saremi,
JHEP {\bf 0610} (2006) 083, 
{\tt[arXiv:hep-th/0601159]}.

\bibitem{bbn}
P. Benincasa, A. Buchel and R. Naryshkin,
Phys. Lett. {\bf B645} (2007) 309, \\
{\tt [arXiv:hep-th/0610145]}.

\bibitem{cvetic}
K. Behrndt, M. Cvetic and W.A. Sabra,
Nucl. Phys. {\bf B553} (1999) 317, \\
{\tt [arXiv:hep-th/9810227]}.

\bibitem{Minwalla}
S. Minwalla,
http://videolectures.net/cern\_minwalla\_nfdg/

\bibitem{yarom}
J. Erdmenger, M. Haack, M. Kaminski and A. Yarom,
JHEP {\bf 0901} (2009) 055, \\
{\tt [arXiv:0809.2488[hep-th]]}.

\bibitem{india}
N. Banerjee, J. Bhattacharya, S. Bhattacharyya, S. Dutta,
R. Loganayagam and P. Surowka,
{\tt [arXiv:0809.2596[hep-th]]}.

\bibitem{son1}
D.T. Son and M.A. Stephanov,
Phys.\ Rev.\ {\bf D77} (2008) 014021, \\
{\tt [arXiv:0710.1084[hep-ph]]}.

\bibitem{Son:2009tf}
D.T. Son and P. Surowka, 
Phys. Rev. Lett. {\bf 103} (2009) 191601, \\
{\tt [arXiv:0906.5044[hep-th]]}.

\bibitem{Yee:2009vw}
H.-U. Yee,
JHEP {\bf 0911} (2009) 085, 
{\tt [arXiv:0908.4189[hep-th]]}.

\bibitem{Torabian:2009qk}
M. Torabian and H.-U. Yee,
JHEP {\bf 0908} (2009) 020, 
{\tt[arXiv:0903.4894[hep-th]]}.

\bibitem{cvetic2}
M. Cvetic and S.S. Gubser,
JHEP {\bf 9904} (1999) 024, 
{\tt [arXiv:hep-th/9902195]}.

\bibitem{ki}
H. Kodama and A. Ishibashi,
Prog.\ Theor.\ Phys.\ {\bf 111} (2004) 29, \\
{\tt [arXiv:hep-th/0308128]}.

\bibitem{Son:2007vk}
D.T. Son and A.O. Starinets,
Ann.\ Rev.\ Nucl.\ Part.\ Sci.\ {\bf 57} (2007) 95, \\
{\tt [arXiv:0704.0240[hep-th]]}.

\bibitem{Sahoo:2009yq}
B. Sahoo and H.-U. Yee,
{\tt [arXiv:0910.5915[hep-th]]}.

\bibitem{bk}
V. Balasubramanian and P. Kraus,
Commun. Math. Phys. {\bf 208} (1999) 413, \\
{\tt[arXiv:hep-th/9902121]}.

\bibitem{kharzeev}
D.E. Kharzeev and H.J. Warringa,
Phys.\ Rev.\  {\bf D80} (2009) 034028, \\
{\tt [arXiv:0907.5007[hep-ph]]}.

\bibitem{Witten:1998xy}
E. Witten,
JHEP {\bf 9807} (1998) 006, 
{\tt [arXiv:hep-th/9805112]}.

\end{thebibliography}
\end{document}